\documentclass[twocolumn,aps,prb,amsmath,amssymb,floatfix,raggedbottom]{revtex4-1}

\pdfoutput=1
\usepackage{graphicx,mathtools,bm,color}

\begin{document}
	
\title{Helical liquid in carbon nanotubes wrapped with DNA molecules}

\begin{abstract}
The measured electric resistance of carbon nanotubes wrapped with DNA molecules depends strongly on the spin of the injected electrons.  Motivated by these experiments,  we study the effect of  helix-shaped potentials on the electronic spectrum of carbon nanotubes. We find that in combination with the curvature-induced spin-orbit coupling inherent  to nanotubes, such a perturbation opens helicity-dependent gaps. Within these partial gaps,  left-moving electrons carry a fixed spin-projection that is reversed for right-moving electrons, and the probability of electrons to transfer through the nanotube correlates  with their helicity. We explain the origin of this effect and show that it can alternatively be induced by twisting the nanotube. Our findings suggest that carbon nanotubes hold great potential for implementing spin filters and may form an  ideal platform to study the physical properties of one-dimensional helical liquids.

\end{abstract}

\author{Yotam Perlitz}
\affiliation{Department of Condensed Matter Physics, Weizmann Institute of Science, Rehovot 76100, Israel}
\author{Karen Michaeli}
\affiliation{Department of Condensed Matter Physics, Weizmann Institute of Science, Rehovot 76100, Israel}

\maketitle

Carbon nanotubes (CNTs), being rolled-up sheets of two-dimensional graphene, inherit many of its  remarkable mechanical and electronic properties.~\cite{Roche2007,Kouwenhoven2015,DresselhausRMP} 
Due to the strong bonds between carbon atoms,~\cite{Gibson1996,Wong1997} the diameter of CNTs can be as small as a few nanometers.~\cite{Toshinari1993} Moreover, high-quality CNTs with long electronic mean-free paths~\cite{Peng2014} can be fabricated by standard techniques~\cite{Toshinari1993,Thess1996,Fischer1997} and are thus readily available. Despite having been around for several decades,  they continue to inspire fundamental research in low-dimensional physics as well as concrete applications in the context of nano electronics.~\cite{Peng2014,Avouris2006}  A unique aspect of CNTs is the enhancement of the extremely weak spin-orbit coupling (SOC) of  graphene due to orbital mixing introduced by the curvature.~\cite{Ando2000,Egger2002,Guinea2006,McEuen2008,Chico2012,Kouwenhoven2013} This SOC has been demonstrated to permit spin manipulation by electric fields.~\cite{McEuen2008,Kuemmet2010} Still, the spin-coherence length in these clean systems is  rather long.~\cite{Hiroki1999} The combination of these two features raises the prospect of using CNTs for spintronics devices, and numerous ideas for such applications have been proposed.~\cite{Belzig2006,Naber2007}

A recent experiment~\cite{Pramanik2015,Pramanik2017} found that CNTs wrapped with DNA molecules perform as strong spin filters.  In particular,  the current-voltage characteristics of wrapped semiconducting CNTs exhibit a strong spin dependence  just above the gap. At low voltage and temperature, spin-polarization of above $60\%$ was measured. This effect becomes weaker as voltage and/or temperature increase; the spin-dependent transport vanishes altogether  for temperatures above $40K$. Similarly strong  spin-selective transport has also been observed in DNA as well as numerous other chiral proteins in the absence of CNTs (for  a recent  review see Ref.~\onlinecite{NaamanReview}). There, spin filtering persists over a large energy range and up to room temperature when electrons pass through insulating organic molecules with  well-defined helicity. Different explanations of  this phenomenon~\cite{Mujica2009,Mujica2012,Gutierrez2012,Eremko2013,Galperin2013,Medina2015,Michaeli2015} rely on a combination of SOC and motion along curved trajectories dictated by the helical structure of the molecules. The experimental observation of spin-filtering in CNTs provides access to this phenomenon from a better understood starting point. In addition, it may allow to integrate such nanoscopic spin-filters into solid-state devices and lead to new spintronic applications. Regarding the latter, we note that mixing  CNTs and DNA molecules is commonly used to separate CNTs and the wrapping occurs spontaneously in this process.~\cite{Zheng2003,Dennis2003,Sanchez-Pomales2010}

\begin{figure}[t]
\begin{flushright}\begin{minipage}{0.5\textwidth}  \centering
        \includegraphics[width=1\textwidth]{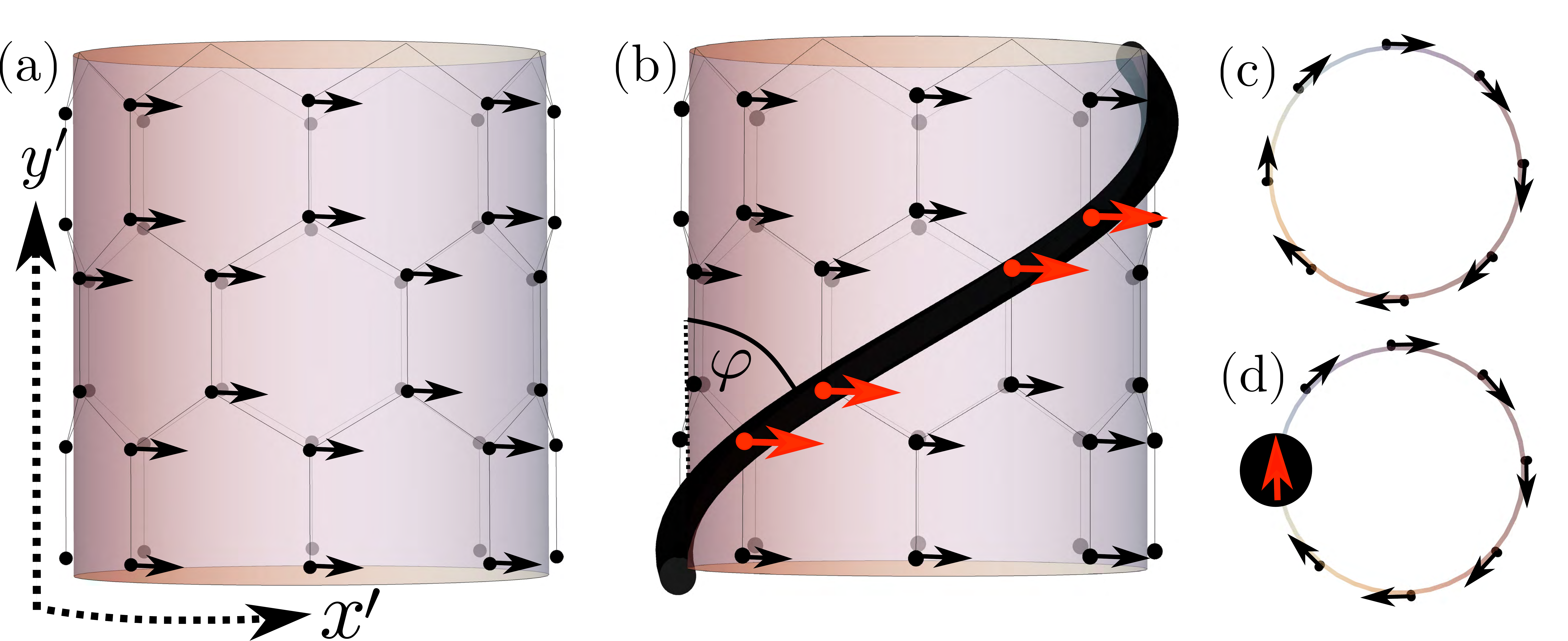} 
                 \caption[0.4\textwidth]{\small Illustration of a zig-zag CNT in the absence (a) and presence (b) of a DNA molecule.  The curvature of the CNT induces SOC that favors alignment of spins in the plane of the cylinder. As explained in the text, alignment along the $y'$-direction does not generate spin-filtering. We, therefore, focus on the component parallel to $x$ and use arrows to indicate the favored spin-alignment for an electron hopping in the counterclockwise direction. Without a molecule,  a closed trajectory around the cylinder   (corresponding to states of fixed angular momentum) leads to cancellation of this type of SOC as illustrated in the top view (c). Such a cancellation is disrupted by the DNA, (b) and (d). As a result, a net SOC  remains, and creates an effective magnetic field that spirals around the CNT (denoted by red arrows). The sign of this magnetic field depends on the angular momentum of the electron.} \label{figure1}
\end{minipage}\end{flushright}
\end{figure}

In this work we analyze how the helix-shaped potential created by a DNA molecule affects the electronic states of  CNTs. We show that in combination with the SOC intrinsic to CNTs, this potential gives rise to spin-filtering. Depending on  the handedness of the molecule, electronic states of one helicity become gapped in several energy windows. At these energies, transmission through the CNT of electrons with one spin is strongly suppressed compared to the other. The preferred spin changes when the transport direction is switched, or the handedness of the chiral potential reversed.  The majority of previous theoretical studies of CNTs subject to chiral potentials~\cite{Rotkin2007,Mele2008} focused on the pseudospin degree of freedom, rather than on the physical spin. Spin transport in such systems  has been  investigated theoretically  in Ref.~\onlinecite{Diniz2012}, where  transmission through a conducting CNT wrapped with DNA was studied. That work calculated the effect of SOC induced by the molecule, and found  that  spin filtering occurs near neutrality point. As we will show later, a helicity-dependent gap does not open at zero energy, and thus, the spin-dependent transport reported in Ref.~\onlinecite{Diniz2012} is not a result of bulk properties.

In the absence of SOC, the spectrum of a CNT is four-fold degenerate and electronic states can be characterized by their spin $s$ and angular momentum $\ell$   parallel to the cylinder axis. The latter corresponds semiclassically to an electron encircling the CNT in the clockwise ($\ell>0$) or counter-clockwise ($\ell<0$) direction. The curvature-induced SOC has a contribution  that favors alignment of spin parallel to the electron trajectory around the CNT,~\cite{Ando2000,Guinea2006} as illustrated in Fig.~\ref{figure1}.  This contribution averages to zero over a full trajectory, and does not affect the energy spectrum. (In addition, there is a contribution favoring alignment of the spin along $\ell$, which lifts the four-fold degeneracy of the electronic spectrum, but does not open a gap.~\cite{McEuen2008}) This cancellation is destroyed when the trajectory is perturbed locally, e.g., by the potential created by the molecule. As a result, electrons experience an excess  magnetic field  tangential to the cylinder at the points of perturbation.  The sign of this effective magnetic field is determined by the circulation direction, corresponding to  $\ell$, thus preserving time reversal symmetry. Its direction spirals around the CNT following the helical wrapping of the DNA molecule. The effect of such an effective magnetic field has been analyzed in one-dimensional wires~\cite{Braunecker2010} (and in the context of chiral molecules~\cite{Eremko2013,Michaeli2015}), where it has been shown to open a gap for states of one helicity, depending on the  rotation direction of the field. This is because such a spiraling magnetic field entangles the spin with \textit{both} linear and angular momenta.

Our fully quantum treatment of the electronic spectrum and spin-selective transport in DNA-wrapped CNTs, which forms the bulk of this paper, is in complete agreement with the semiclassical description. These results also follow directly from symmetry considerations:~\cite{Barros2006} Rolled-up graphene sheets (chiral as well as achiral) are invariant under $\pi$-rotations around an axis perpendicular to $\hat{y}'$. This symmetry prohibits any couplings between spin and momentum parallel to the cylinder axis, which are required for helicity-dependent gaps. The external chiral potential generated by the DNA molecule breaks this $\pi$-rotation invariance. Alternative routes to creating a helical perturbations are by twisting CNTs as discussed below, or by straining chiral CNTs to created unequal hopping amplitudes along different bonds $\boldsymbol{\delta}_i$ (see Fig.~\ref{figure2}). In fact, the latter occurs naturally in  real chiral CNTs where bond lengths are non-uniform,  unlike the hypothetical case of a rolled-up graphene sheet. However, this is a weak effect and $\pi$-rotation remains an approximate symmetry. Consequently, helicity-dependent gaps should be very small, and have not been observed so far. For the rest of this paper, we therefore neglect this weak effect (i.e., we assume equal bonds) and concentrate on the spin-dependence induced by the wrapped molecule. Furthermore,  for specificity, we mostly restrict our discussion to zig-zag CNTs, which can be either metallic or semiconducting.

\section{Model}

\begin{figure}[t]
\begin{flushright}\begin{minipage}{0.5\textwidth}  \centering
        \includegraphics[width=1\textwidth]{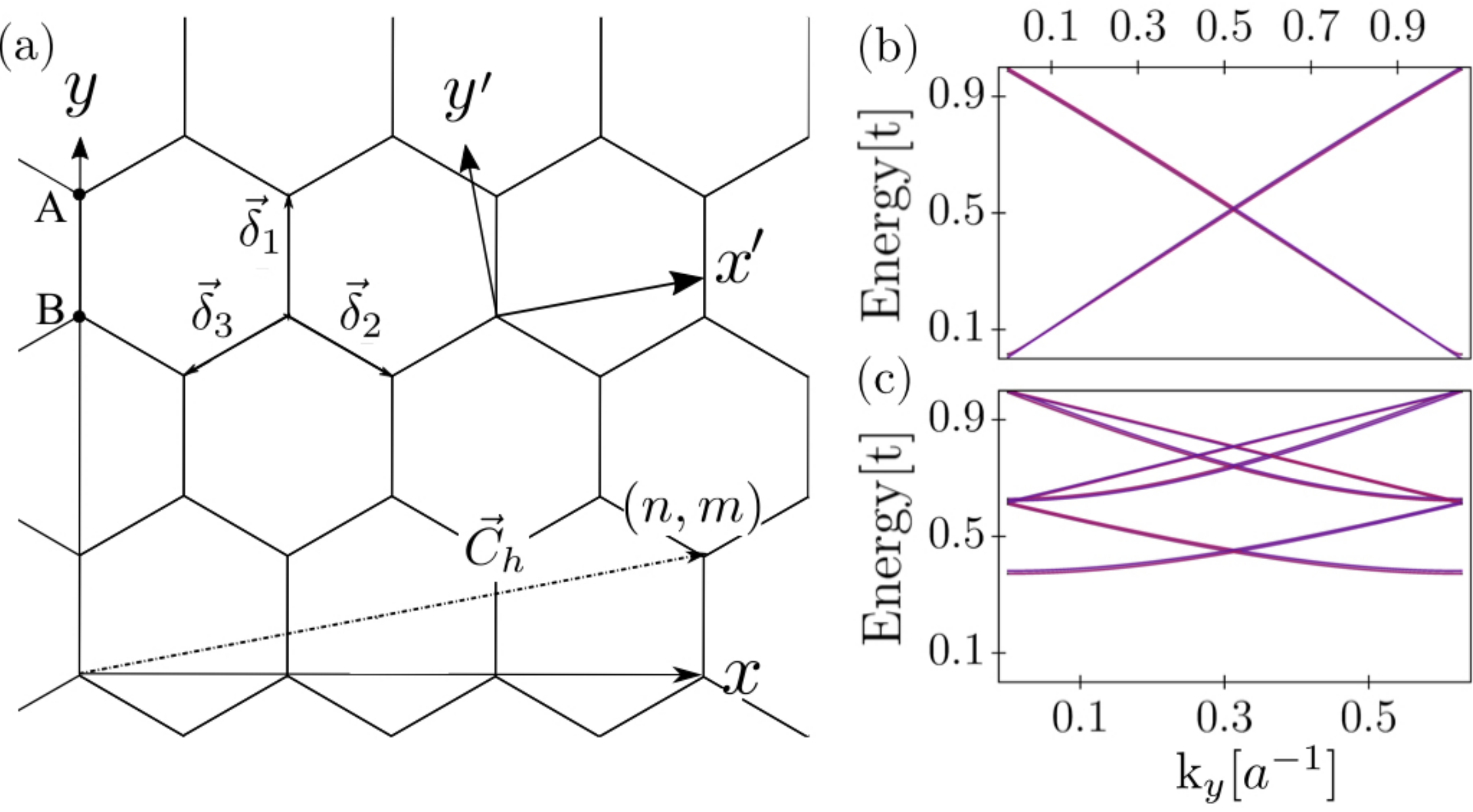} 
                 \caption[0.4\textwidth]{\small  (a) The lattice structure of graphene, which forms a CNT when rolled-up. The vectors $\boldsymbol{\delta}_{i=1,2,3}$ connect nearest-neighbor sites, as illustrated. The coordinate system $(x,y)$ is fixed to the lattice, while $\hat{x}'$ is the folding direction of the CNT. For zig-zag CNTs $\hat{x}'=\hat{x}$, for armchair $\hat{x}'=\hat{y}$, and for a generic chiral folding $\hat{x}'=\hat{C}_h$. The energy spectra of metallic, $N=3$, and semiconducting, $N=5$,  zig-zag CNTs according to Eqs.~(\ref{eq:Hopping H}-\ref{eq:Sx}) are plotted in panels (b) and (c), respectively.  The SOC weakly splits the bands but, crucially, does not open gaps. Note that, in preparation for our subsequent introduction of the molecule, we already plot the energy spectra in the reduced (folded) Brillouin zone.}
                  \label{figure2}
\end{minipage}\end{flushright}
\end{figure}

To study the effect of wrapping CNTs with DNA molecules, we first introduce the Hamiltonian of a bare CNT, $\mathcal{H} = \mathcal{H}_0 + \mathcal{H}_{\text{SOC}}$. The term $\mathcal{H}_0$  describes hopping on a honeycomb lattice, such as in graphene (see Fig.~\ref{figure2}) 
\begin{align} \label{eq:Hopping H}
\mathcal{H}_{0} &= \sum_{\boldsymbol{r},i,s}  t_{i} A^{\dagger}_{\boldsymbol{r}+ \boldsymbol{\delta}_i,s} B_{\boldsymbol{r},s} + \text{h.c.}
\end{align}
Here $A^{\dagger}_{\boldsymbol{r},s}$ ($B_{\boldsymbol{r},s}^{\dagger}$)  creates a spin-$s$ electron in the $\pi$ ($p_z$) orbital on the A (B) sub-lattice; the vector $\mathbf{r}=(x,y)$ labels  unit cells as illustrated in Fig.~\ref{figure2}. The single-orbital Hamiltonian $\mathcal{H}_0$ captures the low-energy properties of undoped CNTs whose chemical potential lies near the center of the $\pi$ band. Additional bands that are comprised of $\sigma$  ($s$, $p_x$ and $p_y$) orbitals are found at energies $\gtrsim2eV$ away from the chemical potential. To transition from a graphene sheet to a CNT, the former needs to be rolled up, which amounts to imposing periodic boundary conditions~\cite{Saito1992} along the $x'$-axis (around the cylinder).  It is convenient to chose the spin quantization axis parallel to the central axis of the CNT, i.e., along $\hat{y}'$.   $t_i$  is the  hopping amplitude between nearest neighbors connected by the vectors $\boldsymbol{\delta}_{i}$   with  $ \boldsymbol{\delta}_{1} = \hat{y}a/\sqrt{3}$, $ \boldsymbol{\delta}_{2} = \hat{x}a/2-\hat{y}a/\sqrt{3} $ and $ \boldsymbol{\delta}_{3} = -\hat{x}a/2-\hat{y}a/\sqrt{3} $, cf. Fig.~\ref{figure2}.

In graphene, SOC is negligible due to the combination of a large energy difference between the $\pi$ and $\sigma$ orbitals and strictly zero hopping amplitude between them.  The curvature of CNTs allows hopping between $\pi$ and $\sigma$ orbitals and  gives rise to a  significantly larger SOC.~\cite{Ando2000,Guinea2006} This curvature-induced SOC only affects hopping along the curved direction, i.e., bonds with $\boldsymbol{\delta}_i\cdot\hat{x}'\neq0$, and takes the form $ \mathcal{H}_{\text{so}} =  \mathcal{H}_{\text{so}_{1}} + \mathcal{H}_{\text{so}_{2}}$. The first term modifies the hopping amplitude depending on the spin projection along the $y'$-axis~\cite{Ando2000,Guinea2006} 	
\begin{align} \label{eq:Sy}
\mathcal{H}_{\text{so}_{1}}&=\Gamma_1\sum_{\mathbf{r},i,s}  s (\hat{\delta}_i\cdot\hat{x}')  A_{\boldsymbol{r}+\boldsymbol{\delta}_i,s}^{\dagger} B_{\boldsymbol{r},s}+\text{h.c.}
\end{align}
Here  $ \Gamma_1 \approx 0.02a\Delta_{\text{ASOC}}/R  $  where $ R $ is the CNT radius and $ \Delta_{\text{ASOC}}\simeq10 $meV  is the atomic SOC of carbon.~\cite{NIST2005}
The second contribution to the SOC combines hopping with a spin flip~\cite{Ando2000,Guinea2006} 
\begin{align} \label{eq:Sx}
	\mathcal{H}_{\text{so}_{2}}=
	& \Gamma_2
\sum_{\boldsymbol{r},i,s} (\hat{\delta}_i\cdot\hat{x}')^{2}(\hat{\delta}_i\cdot\hat{y}')\\\nonumber&\times\left[e^{-is(\theta_{\boldsymbol{r}}+\theta_{\boldsymbol{r}+\boldsymbol{\delta}_i})/{2}} A_{\boldsymbol{r}+\boldsymbol{\delta}_i,s}^{\dagger}B_{\boldsymbol{r},-s}+\text{h.c.}\right]\hspace{0.5mm},
\end{align}
where $ \Gamma_2 \approx 2a\Delta_{\text{ASOC}}/R$. The position-dependent phase $e^{-is(\theta_{\boldsymbol{r}}+\theta_{\boldsymbol{r}+\boldsymbol{\delta}_i})/2}$ reflects the fact that $\mathcal{H}_{\text{so}_2}$ favors spin alignment  tangential to the cylinder. The argument  $\theta_{\boldsymbol{r}}=\boldsymbol{r}\cdot\hat{x}'/R$ is the angle of a vector tangent to the cylinder at position $\boldsymbol{r}$, as  illustrated in Fig.~\ref{figure1}. This term averages to zero in the absence of a chiral potential and does not affect the energy spectrum.

The DNA molecule wraps around the CNT due to interactions between the $\pi$ orbitals of both systems ($\pi -\pi$ stacking). These interactions give rise to a stable  CNT-DNA hybrid  with helical wrapping that is typically  found  to be commensurate with the hexagonal lattice.~\cite{Zheng2003,Dennis2003,Yarotski2009} 
There are two natural ways to encode the influence of the molecule on the low-energy, single orbital Hamiltonian of the CNT: (i) By modifying the hopping amplitudes. (ii) By adding on-site potentials. Because of screening, in both cases only sites directly underneath the molecule are significantly affected. Here we adopt (i) for which the hopping parameter in Eq.~\ref{eq:Hopping H} and the SOC amplitudes in Eqs.~\ref{eq:Sy} and~\ref{eq:Sx}   become  space dependent, $t_{i}\rightarrow t_{i}(\mathbf{r}')$ and $\Gamma_{x,y}^{i}\rightarrow\Gamma_{x,y}^{i}(\mathbf{r}')$. This choice of implementing the potential  readily separates into spin-dependent and independent contributions. In appendix~\ref{appendixA} we show that (ii), which does not allow this kind of separation, results in a similar electron spectrum and spin-dependent transport. Both of these prescriptions to incorporate the effect of the molecule take into account only the direct modifications of the lowest conduction band. In particular, they do not involve changes to the hybridization between $\pi$ and $\sigma$ orbitals generated by the molecule; the position-dependent SOC terms are caused by local changes to all hopping amplitude, independent of spin. Including such a molecule-induced hybridization would result in an additional SOC contribution of the Rashba-type.~\cite{Guinea2006,Diniz2012} As we argue below and show explicitly in appendix~\ref{appendixB}, this additional term does not qualitatively affects our results.

\section{Energy spectrum}

To understand the effect of the molecule on the electron spin, we analyze the position-dependent SOC,  $\Gamma_{1,2}^i(\mathbf{r})$,  while keeping a uniform hopping parameter $t_i$  in Eq.~\ref{eq:Hopping H}.  In the absence of SOC, the energy spectrum of the CNT described by $\mathcal{H}_0$ is $E=\pm\sqrt{1+4\cos(k_xa/2)\cos(\sqrt{3}k_ya/2)+4\cos^2(k_xa/2)}$. 
For specificity, we assume a zig-zag CNT for which the $x$ and $x'$ axes coincide, and the periodic boundary conditions take the form $k_x=2\pi \ell/Na$, where $\ell=1,2,\ldots,2N$ and $Na$ is the CNT circumference. The SOC terms in momentum space read $\mathcal{H}_{\text{so}_{1},\text{so}_{2}}={h}_{\text{so}_{1},\text{so}_{2}}^{s,s'}(\ell,\ell';k_y,k_y') A_{\ell,k_y,s}^{\dagger} B_{\ell',k_y',s'}+\text{h.c}$  (repeated indices are summed over), where 
\begin{align} \label{eq:SOCmom}
h_{\text{so}_{1}}\hspace{-1mm}\propto &s e^{i\frac{k_ya}{2\sqrt{3}}}\sin\left[\frac{\pi \ell}{N}\right]\delta_{k_y,k_y'} \delta_{s,s'}\left( \delta_{\ell,\ell'} + \delta_{2N-\ell,\ell'} \right)\hspace{0.5mm},\\\nonumber
h_{\text{so}_{2}}\hspace{-1mm}\propto &e^{i\frac{k_ya}{2\sqrt{3}}}\hspace{-0.5mm}\cos\hspace{-1mm}\left[\frac{\pi(2\ell+s)}{2N}\right]\hspace{-1.mm} \delta_{k_y,k_y'} \delta_{-s,s'}\\\nonumber&\times\left( \delta_{\ell+s,\ell'} + \delta_{2N-\ell+s,\ell'} \right)\hspace{0.5mm}. 
\end{align}
The first term, $h_{\text{so}_{1}}$, is diagonal in spin and momentum and thus results in a trivial energy shift. In contrast, $h_{\text{so}_{2}}$ contains off-diagonal terms which mix different spins and different $x$-momenta ($\ell$) and can in principle open gaps.

\begin{figure}
\includegraphics[width=\linewidth]{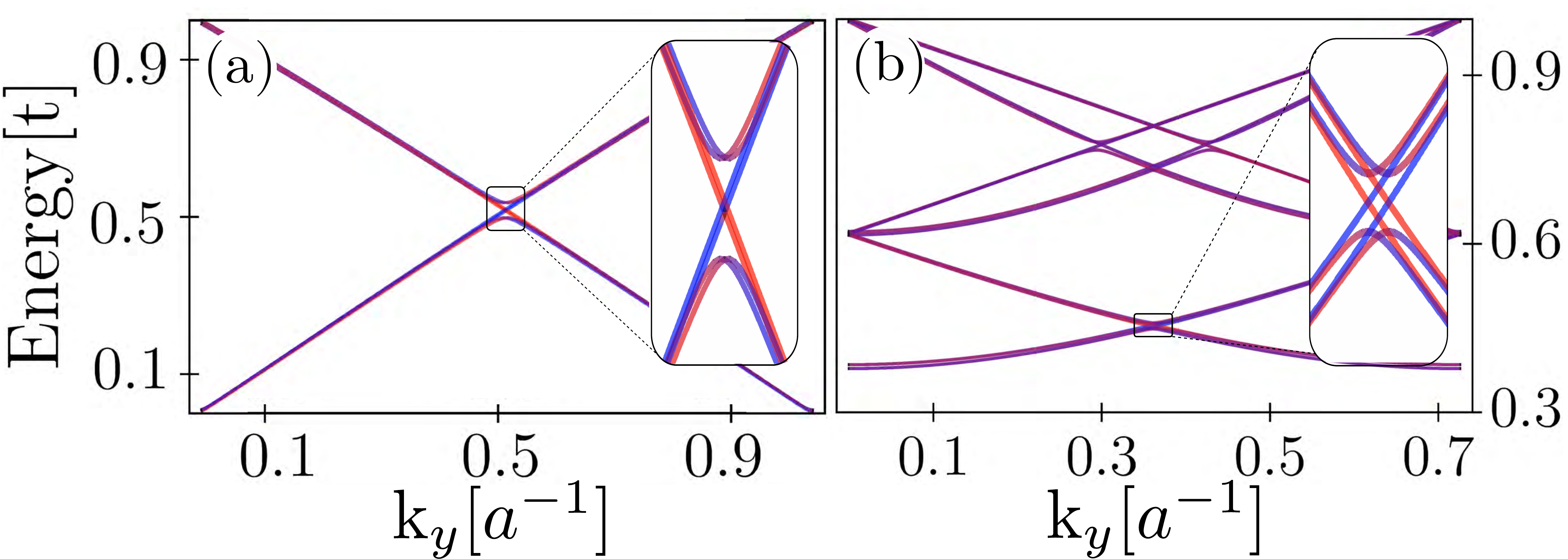}
\caption{ The energy spectra of  a metallic $N=3$ CNT (a) and a semiconducting $N=5$ CNT (b),  each wrapped by a DNA molecule at angle $ \varphi= \pi /6 $. The spin-projection along the cylinder axis is indicated by blue (positive), red (negative) or purple (no net projection). The zoom-ins around the gaps clearly show the well-defined helicity of the remaining states. 
We note that for the metallic CNT, no gap occurs at zero energy. }
\label{figure3}
	\end{figure}

To examine  whether the SOC indeed opens  gaps, it is convenient to express $\mathcal{H}_{\text{so}}$ in the eigenbasis of $\mathcal{H}_{0}$, i.e., $\psi_{\ell,k_y,s}^{\eta}$, where $\eta=\pm$ denotes positive and negative energy states. Since $h_{\text{so}_2}$ intertwines the spin direction with the angular momenta $n$ and $n'$, the Hamiltonian couples the $(\ell,\uparrow)$ and $(\ell+1,\downarrow)$ states but not $(\ell,\downarrow)$ and $(\ell+1,\uparrow)$. [The same holds for the coupling between $(\ell,\uparrow)$ and $(2N-\ell+1,\downarrow)$.] Therefore, if a gap was to open for the former two, the remaining states would be helical.  In the absence of SOC, the bands with quantum numbers $(\ell,\uparrow)$ and $(\ell',\downarrow)$  cross at $k_y=k_0$, with $\cos[\sqrt{3}k_0/2]=\cos[\pi \ell/N]+\cos[\pi \ell'/N]$.  Setting $k_y=k_0+q$ and expanding the $2\times2$  Hamiltonian $\mathcal{H}_0+\mathcal{H}_{\text{so}}=\Psi_{\ell,\ell',\eta}^{\dagger}(q)h_{\ell,\ell'}^{\eta}(q)\Psi_{\ell,\ell',\eta}(q)\delta_{k_y,k_y'}(\delta_{\ell',\ell+1}+\delta_{\ell',2N-\ell+1})$ to leading order in $q$, we find 
\begin{align} \label{eq:Linearized1}
h_{\ell,\ell'}^{\eta}(q)\hspace{-1mm} \approx\hspace{-1mm} 
 \left[\begin{matrix} E_0\hspace{-0.5mm}-\hspace{-0.5mm}v_1q\hspace{-0.5mm}+\hspace{-0.5mm}F_{\ell,\ell'}\hspace{-0.5mm}(q)&  G_{\ell,\ell'}\hspace{-0.5mm}(q) \\   
			G_{\ell,\ell'}^{*}\hspace{-0.5mm}(q) & E_0\hspace{-0.5mm}+\hspace{-0.5mm}v_2q\hspace{-0.5mm}+\hspace{-0.5mm}F_{\ell,\ell'}\hspace{-0.5mm}(q)
			\end{matrix}\right]\hspace{-0.5mm}.
\end{align}
Here,  $E_0$ is the energy at the crossing point, $v_{1,2}=\sqrt{3}t^2\left[2\tan(\pi/2N)\sin^2(\sqrt{3}k_0/2)\pm\sin(\sqrt{3}k_0)\right]/4E_0$,  and $ \Psi_{\ell,\ell',\eta}^T(q)=[\psi_{\ell,k_0+q,\uparrow}^{\eta},\psi_{\ell',k_0+q,\downarrow}^{\eta}]$. Crucially, the off-diagonal term  $G_{\ell,\ell'}(q)= {\cal O} (q)$  vanishes at the band crossing. The diagonal term $F_{\ell,\ell'}(q)=\tan(\pi/2N)\sin[\sqrt{3}k_0/2]+ {\cal O}(q)$ merely shifts the energy. This zero reflects the fact that upon averaging over the circumference of the CNT, the effect of $\mathcal{H}_{\text{so}_2}$ disappears. Moreover, since the SOC mixes states with different angular momenta it cannot open a gap at the neutrality point where $\psi_{\ell,k_y,s}^{+}$ and $\psi_{\ell,k_y,s}^{-}$ cross. From the above derivation it follows that the SOC term alone cannot open a gap (see also Fig.~\ref{figure2} for the spectra of two zig-zag CNTs with $\Gamma_1=3\cdot10^{-4}t/N$ and $\Gamma_2=3\cdot10^{-2}t/N$). This result equally applies for chiral CNTs, where the periodic boundary condition in the $x'$-direction results in a modified quantization condition for a linear combination of $k_x$ and $k_y$. Still, the SOC only couples $\psi_{\ell,k_y',\uparrow}^{+}$ with $\psi_{\ell+1,k_y',\downarrow}^{+}$, where $k_y'$ is orthogonal to  $k_x'$.
   
	\begin{figure}
		\includegraphics[width=\linewidth]{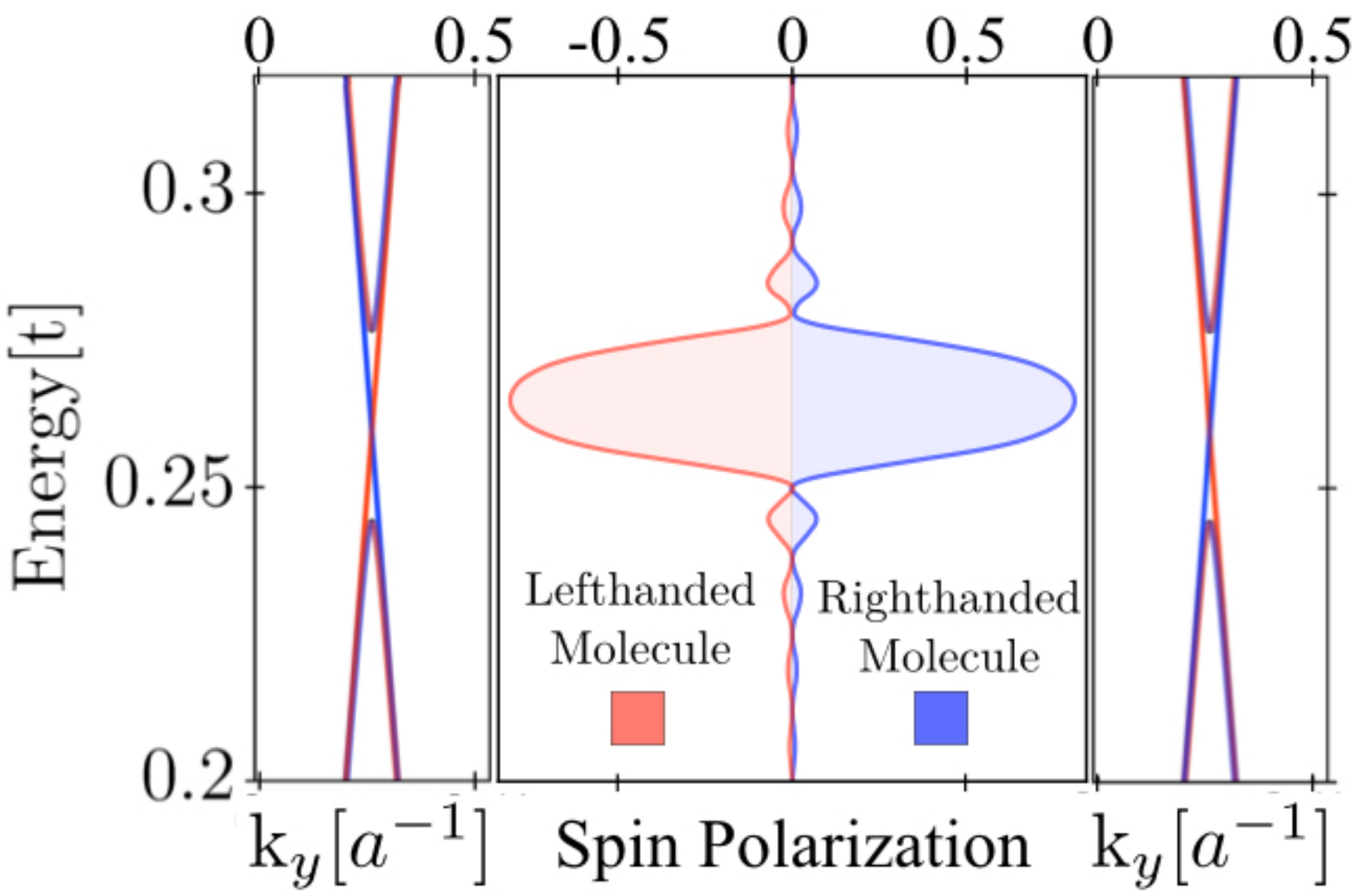}
		\caption{ Spin-selective electron transmission through a metallic $N=6$ CNT wrapped with a right- or a left-handed molecule, $  \varphi=\pm \pi/6$. The  spin polarization $P = \left(T_{\uparrow}-T_{\downarrow}\right) / \left(T_{\uparrow}+T_{\downarrow}\right)$ is computed from the transmission probabilities $T_{s}$ for electrons of spin $s$. It is substantial at energies where the partial gap opens (shown on the sides). The two cases, a right- and a left-handed molecule, exhibit spin polarization with equal magnitude but opposite sign.  }
		\label{figure4}
	\end{figure}

We now turn to analyze the effects of a  DNA molecule. The SOC terms acquire additional contributions that are non-zero only along  lines defined by $y=(x+Nam)\tan{\varphi}$ where $m \in \mathbb{Z}$, and the wrapping angle $\varphi\in[-\pi/2,\pi/2]$ is positive (negative) for right (left) handed molecules.  As a result, the SOC terms in  Eq.~\ref{eq:SOCmom} are no longer diagonal in the momentum along the $y$-direction.  For example, taking the wrapping angle to be $\varphi=\pi/6$, the  additional contributions to the SOC are   
\begin{align} \label{eq:SOCmomMol}
&\delta{h}_{\text{so}_1}\propto se^{i\frac{k_ya}{2\sqrt{3}}}\sin\left[\frac{\pi \ell}{N}\right]\delta_{k_y+Q(\ell-\ell'),k_y'} \delta_{s,s'}\hspace{0.5mm},\\\nonumber
&\delta{h}_{\text{so}_2}\propto  e^{i\frac{k_ya}{2\sqrt{3}}}\cos\hspace{-1mm}\left[\frac{\pi(2\ell+s)}{2N}\right]  \delta_{k_y+Q(\ell+s-\ell'),k_y'} \delta_{-s,s'}\hspace{0.5mm},
\end{align}
where  $Q(x)=2\pi x/\sqrt{3}Na$. The broken translation symmetry suggest using a reduced Brillouin zone, with bands labeled by $n$ in addition to $\ell$, $s$ and the reduced momentum $k_y^{\text{red}}$ where $k_y=k_y^{\text{red}}+Q(n)$. Bands that cross in the reduced  Brillouin zone couple when satisfying the  delta-function involving  $Q$ in Eq.~\ref{eq:SOCmomMol}. Thus, as before, the states $(\ell,\uparrow)$ and $(\ell+1-n,\downarrow)$  mix but not $(\ell,\downarrow)$ and  $(\ell+1-n,\uparrow)$. The $2\times2$  Hamiltonian in the vicinity of these energies is  $\mathcal{H}=\Psi_{\ell,\ell',\eta}^{\dagger}(q)h_{\ell,\ell'}^{\eta}(q)\Psi_{\ell,\ell',\eta}(q)\delta_{k_y-Q(n),k_y'}\delta_{\ell',\ell+s-n}$, where $h_{\ell,\ell'}^{\eta}(q)$ has the same structure as in Eq.~\ref{eq:Linearized1}. The main differences between this Hamiltonian and the case  without a molecule (discussed above) is in the definition of the states $\Psi_{\ell,\ell',\eta}^T=[\psi_{\ell,k_0+q,\uparrow}^{\eta},\psi_{\ell',k_0+q-Q(\ell+s-\ell'),\downarrow}^{\eta}]$ and   $G_{\ell,\ell'}(q\rightarrow0)\neq 0$.

 	\begin{figure}
		\includegraphics[width=\linewidth]{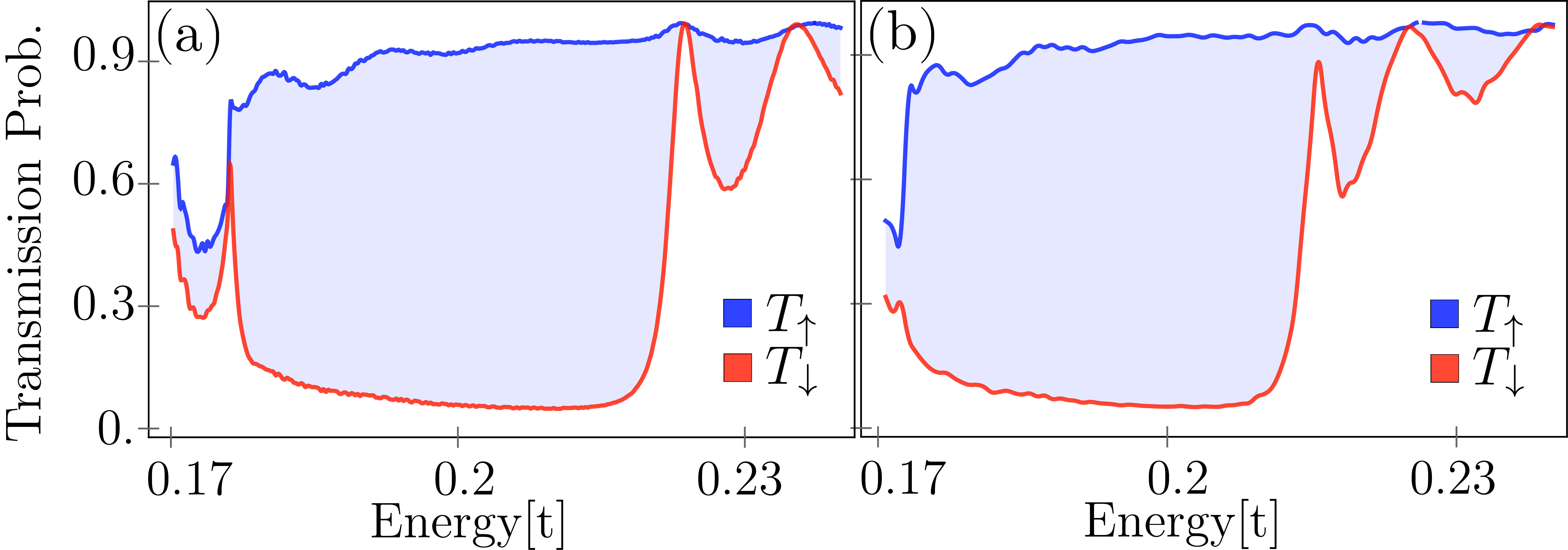}
		\caption{		Spin-dependent transmission through clean (a) and weakly disordered (b) semiconducting $N=11$  CNTs of length $154a/\sqrt{3}$ wrapped with DNA at $\varphi=\pi/6 $. The transmission is significant only above the semiconducting gap $\approx0.17t$. In both cases, transmission of $s=~\downarrow$  electrons is strongly suppressed compared to $s=~\uparrow$ at energies within the partial gap. In the clean case (a) the spin polarization diminishes at below $\approx0.18t$. In contrast, in the weakly disordered case (b) significant spin polarization persists to the lowest energies above the band gap.} 
		\label{figure5}
	\end{figure}

The  spectrum of the full Hamiltonian, Eqs.~\ref{eq:Hopping H}-\ref{eq:Sx}, in the presence of the molecule (encoded in the  position-dependent  SOC) is presented in Fig.~\ref{figure3}. There we show the energy spectra for metallic ($N=3$) and semiconducting ($N=5$) zig-zag CNTs wrapped with a DNA molecule at  $\varphi=\pi/6$. The couplings are taken to be $\Gamma_1=3\cdot10^{-4}t/N$ and $\Gamma_2=3\cdot10^{-2}t/N$ on all sites unaffected by the molecule (corresponding to $\Delta_{\text{ASOC}}\approx5\cdot{10}^{-3}t$).  On sites underneath the molecule, we took these parameters to be larger by a factor of 10.   The relevant parameters for experimental systems are discussed below. The opening of helicity-dependent gaps of size $0.03t$ (metallic) and $0.01t$ (semiconducting) is clearly visible in Fig.~\ref{figure3}---for comparison, see Fig.~\ref{figure2} for the spectra in the absence of DNA molecules. Within the energies of the partial gap, only states of one helicity remain. We note that no helicity-dependent gaps open near the neutrality point of the metallic CNT.

\section{Spin dependent transmission}

To study the spin-filtering properties of  CNTs wrapped with DNA, we numerically calculate the transmission probability per spin, $T_{s}$, as a function of energy. For this purpose, we assume the molecule wraps only a segment of the CNT, and let the uncovered parts acts as electrodes. In Fig.~\ref{figure4} we show the spin-polarization, $P=(T_{\uparrow}-T_{\downarrow})/(T_{\uparrow}+T_{\downarrow})$, for electron transport through a metallic CNT ($N=6$) as a function of energy. A clear window with spin polarization of more than $90\%$ occurs at the energies where helicity dependent gaps open.  For the transport calculation, we took large SOC parameters, $\Gamma_1=10^{-3}t$ and $\Gamma_2=10^{-1}t$ on sites directly below the molecule and $10$ times smaller elsewhere. This choice of parameters allowed us to perform our simulation on relatively short molecules. This is because the spin polarization increases exponentially with length, $P\sim P_{\text{max}} - e^{-r/\xi}$, where $\xi$ is inversely proportional to the partial gap (for the full length dependence see Appendix~\ref{appendixC}).

For metallic CNTs, the first partial gap opens at a relatively high energy. In the experiment,~\cite{Pramanik2015,Pramanik2017} in contrast, semiconducting CNTs have been measured and the largest spin polarization was observed at the  lowest voltage where current flows. The spectrum of a representative semiconducting CNT shows a helicity-dependent gap opening not far from the bottom of the conduction band, see Fig.~\ref{figure3}.  Consequently, a clean semiconducting CNT does not exhibit significant spin filter at the lowest energies within our model. However, the band below the partial gap is relatively narrow and is thus expected to easily localize in the presence of disorder. To test this, we calculated the transmission of such a CNT with a realistic circumference ($N=11$) and weak chemical potential disorder $\delta \mu \in [-0.05t,0.05t]$. The spin-dependent transmissions in the clean and weakly disordered cases are shown in Fig.~\ref{figure5}. The presented data is obtained by averaging $T_s$ over 25 realizations, and  only small sample variations were observed. We find that, unlike the clean case, the disordered one exhibits significant spin-polarization at the lowest energies, similar to the experimental findings.

Our discussion so far has focused on the effect of the molecule within a low-energy single-orbital model. Allowing for hybridization of $\pi$ and $\sigma$ orbitals induced by the molecule potential gives rise to a Rashba-type SOC.~\cite{Guinea2006,Diniz2012} The effect of such a SOC, on spin transport of metallic CNTs wrapped with DNA molecules has been analyzed in Ref.~\onlinecite{Diniz2012}. There,  spin filtering has been found near the neutrality point. As we explicitly show in  appendix~\ref{appendixB}, replacing the curvature induced SOC with the Rashba type gives qualitatively the same result as in our earlier discussion.  This is not surprising since the only difference between the Rashba SOC and Eqs.~\ref{eq:Sy}-\ref{eq:Sx} is that the former does not vanish for hopping parallel to the cylinder axis ($y'$-direction). These additional contributions, however,  do not intertwine the spin and angular momentum. In other words, although they include spin flips  their effect is closer to $h_{\text{so}_1}$  than to $h_{\text{so}_2}$. Consequently, the Rashba SOC does not open any new helicity-dependent gaps, and  does  not give rise to further spin filtering on top of the one discussed above.

	\begin{figure}
		\includegraphics[width=\linewidth]{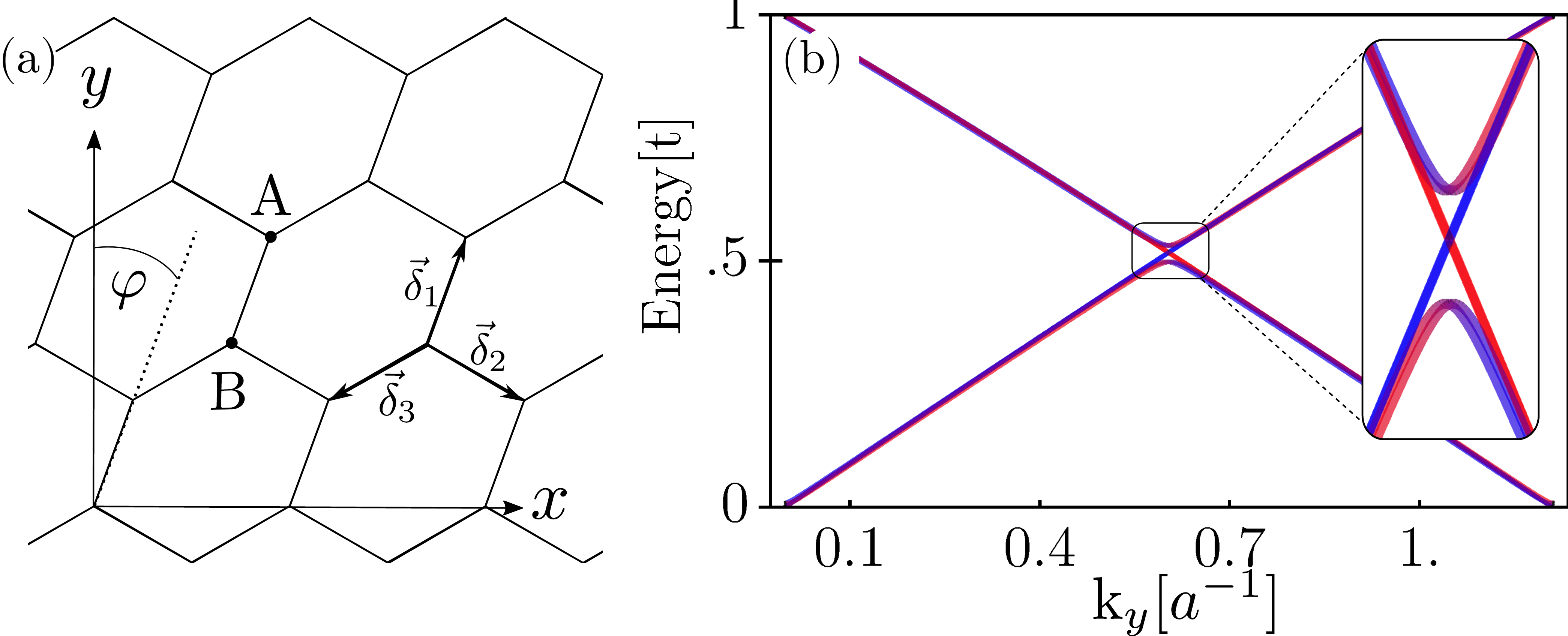}
		\caption{(a) Twisted CNTs can be viewed as rolled-up sheets of distorted graphene. Specifically, we tilt the vector $\boldsymbol{\delta}_1$ by an angle  $\varphi\approx \pi/5$, while $\boldsymbol{\delta}_{2,3}$ are unchanged. The corresponding spectrum for a metallic $N=6$ CNT with $\Gamma_1=2\cdot10^{-1}t$ and $\Gamma_2=2\cdot10^{-3}t$ on all sites is plotted in (b) with a zoom-in on the helicity-dependent gap in the inset. }
		\label{figure6}
	\end{figure}

\section{Connection to experiment}

The above theoretical analysis suggests that the spin-selective transport measured in Refs.~\onlinecite{Pramanik2015,Pramanik2017} results from the partial gaps that open due to the combined effects  of the curvature-induced SOC and of the DNA molecule. To estimate the size of these helicity-dependent gaps within our model,  knowledge of the effective SOC parameters underneath the molecule is required. As shown in appendix~\ref{appendixA},  the effective SOC is not solely determined by the direct change in $\Gamma_{1,2}(\mathbf{r})$;  it also incorporates the electrostatic potential generated by the molecule, as well as the modified hopping amplitudes. Precise values of the parameters can only be determined via ab-initio calculations, but a rough estimate can be obtained as follows. The potential on the carbon sites due to  the molecule was found~\cite{Yarotski2009} to be around $0.1eV$. In appendix~\ref{appendixA} we show that the size of the partial gap induced by a helical potential of strength $0.5t$ is equivalent to the one due to SOC with $\Gamma_2=3\cdot10^{-1}t/N$ underneath the molecule. For  graphene, $t\approx2.7eV$, which yields gaps of size  $1meV-10meV$ in CNTs with radius of a few nanometers. This is consistent with the characteristic energy scale extracted from the observed temperature dependence of spin-polarization in transport, which is substantial up to $40K$. The voltage-dependence of the spin-selective transport is more complicated, and its analysis likely requires including electron-electron interactions.

\section{Twist-induced spin selectivity}

Our analysis of CNTs wrapped with DNA molecules  suggests that any spiral perturbation will lead to similar modifications of the spectrum and give rise to spin-dependent transport. An alternative way would be twisting the CNT. For a zig-zag CNT this can be encoded by rotating  the $\boldsymbol{\delta}_1$ bond  by an angle $\varphi$ (see Fig.~\ref{figure6}), i.e., the lattice coordinates become  $(x,y)\rightarrow (x+y\sin\varphi,y)$ while the basis vector $\boldsymbol{\delta}_2$ remains unaltered. Such a twist modifies the spin-independent hopping $t_1$ as well as the SOC. To focus on spin-related phenomena, we neglect the change in $t_1$  and analyze the SOC given by Eqs.~\ref{eq:Sy} and~\ref{eq:Sx}. The twist renders the angle $\theta_r=x+y\sin\varphi$ in $\mathcal{H}_{\text{so}_2}$ spatially non-uniform (in both directions), and results in non-zero SOC terms on  all bonds. In momentum space $\mathcal{H}_{\text{so}_{1},\text{so}_{2}}={h}_{\text{so}_{1},\text{so}_{2}}^{s,s'}(\ell,\ell';k_y,k_y') A_{\ell,k_y,s}^{\dagger} B_{\ell',k_y',s'}+\text{h.c.}$, with
\begin{align} \label{eq:SOCmomTwist1}
{h}_{\text{so}_1} \propto &~s (\hat{\delta}_i\cdot\hat{x}) e^{-i\mathbf{k}\cdot\boldsymbol{\delta}_i}\delta_{k_y,k_y'} \delta_{s,s'}\left(\delta_{\ell,\ell'}+\delta_{2N-\ell,\ell'}\right)\hspace{0.5mm},
\end{align}
and
\begin{align} \label{eq:SOCmomTwist2}
{h}_{\text{so}_2} \propto &~(\hat{\delta}_i\cdot\hat{x})^2(\hat{\delta}_i\cdot\hat{y}) e^{-i\mathbf{k}\cdot\boldsymbol{\delta}_i-i\pi s(\hat{\delta}_i\cdot\hat{x})/N} \delta_{-s,s'}\\\nonumber
&\times \delta_{k_y+\tilde{Q}(\ell+s-\ell'),k_y'} 
\left(\delta_{\ell+s,\ell'}+\delta_{2N-\ell+s,\ell'}\right)\hspace{0.5mm},
\end{align}
where $\tilde{Q}(x)=2\pi x\sin\varphi/Na$. The outcome of such SOC terms can be analyzed analogously to Eq.~\ref{eq:SOCmomMol}. Specifically,  the  $2\times2$ Hamiltonian near these crossing points is $\mathcal{H}=\Psi_{\ell,\ell',\eta}^{\dagger}h_{\ell,\ell'}^{\eta}\Psi_{\ell,\ell',\eta}\delta_{k_y-\tilde{Q}(n),k_y'}(\delta_{\ell',\ell+s}+\delta_{\ell',2N-\ell+s})$. Here, $h_{\ell,\ell'}^{\eta}(q)$ is given by Eq.~\ref{eq:Linearized1}, and $\Psi_{\ell,\ell',\eta}^T=[\psi_{\ell,k_0+q,\uparrow}^{\eta},\psi_{\ell',k_0+q-\tilde{Q}(\ell+s-\ell'),\downarrow}^{\eta}]$. For small twist angles, $\varphi\ll1$, the crossing points shift by $\delta k_0\propto2\pi(\ell+s-\ell')\varphi/Na$ compared to their position in a bare CNT. There, we saw that $ G_{\ell,\ell'}(q)$ vanished at the crossings. This is no longer the case once these points are shifted, and $G_{\ell,\ell'}(q)\propto\Gamma_2\varphi$. Consequently, partial gaps open as  shown in Fig.~\ref{figure6}. There, a helicity-dependent gap of size $0.03t$ arises for metallic CNT with $N=6$, $\Gamma_1=2\cdot10^{-3}t$, $\Gamma_2=2\cdot10^{-1}t$  and $\varphi=\sin^{-1}(1/\sqrt{3})\approx\pi/5$.  Twisted CNTs with angles up to $\pi/18$ are routinely created in experiments~\cite{Hall2007}, and it would be interesting to look for spin-dependent transport in these systems.  For CNTs  with $R\approx1nm$ and $\Delta_{\text{ASOC}}\approx10 meV$ such a twist would result in partial gaps of order $100\mu eV$.

\section{Conclusion}

We have shown  that helicity-dependent gaps can be created in CNTs via helix-shaped potentials or through creating a (mechanical) twist. Within these partial gaps, all bulk states of left-moving electrons  carry a fixed spin-projection which is reversed for right-moving electrons.   As a result, transport through such CNTs is highly spin-dependent. 
For the case of DNA-wrapped CNTs, a recent experiment has indeed found that the resistance strongly depends of the spin of injected electrons. This  paves a clear path towards   CNT-based spin valves and filters without  magnetic field.~\cite{Michaeli2017} Our results also suggest that CNTs are a natural platform to study various exciting  phenomena that have been predicted to arise in one-dimensional helical systems. For example, by coupling a DNA-wrapped (or twisted) CNT to a quantum dot, one could probe the Kondo effect in a helical liquid.~\cite{Maciejko2009,Altshuler2015} Another very interesting setup would be Josephson junctions, formed by coupling CNTs to superconducting leads. Here, unconventional Andreev bound states are expected to occur which can strongly modify the transport properties compared to a conventional junction.

\textit{\ Acknowledgements:} We thank Binghai Yan and Philip Kim for helpful discussions. This work is supported by the Israel Science Foundation Grant No.~1889/16 and the Minerva Foundation (KM).

\bibliographystyle{apsrev4-1}
\bibliography{ciss}

\appendix

\section{Alternative models}\label{appendixA}

\begin{figure}[b]
		\includegraphics[width=\linewidth]{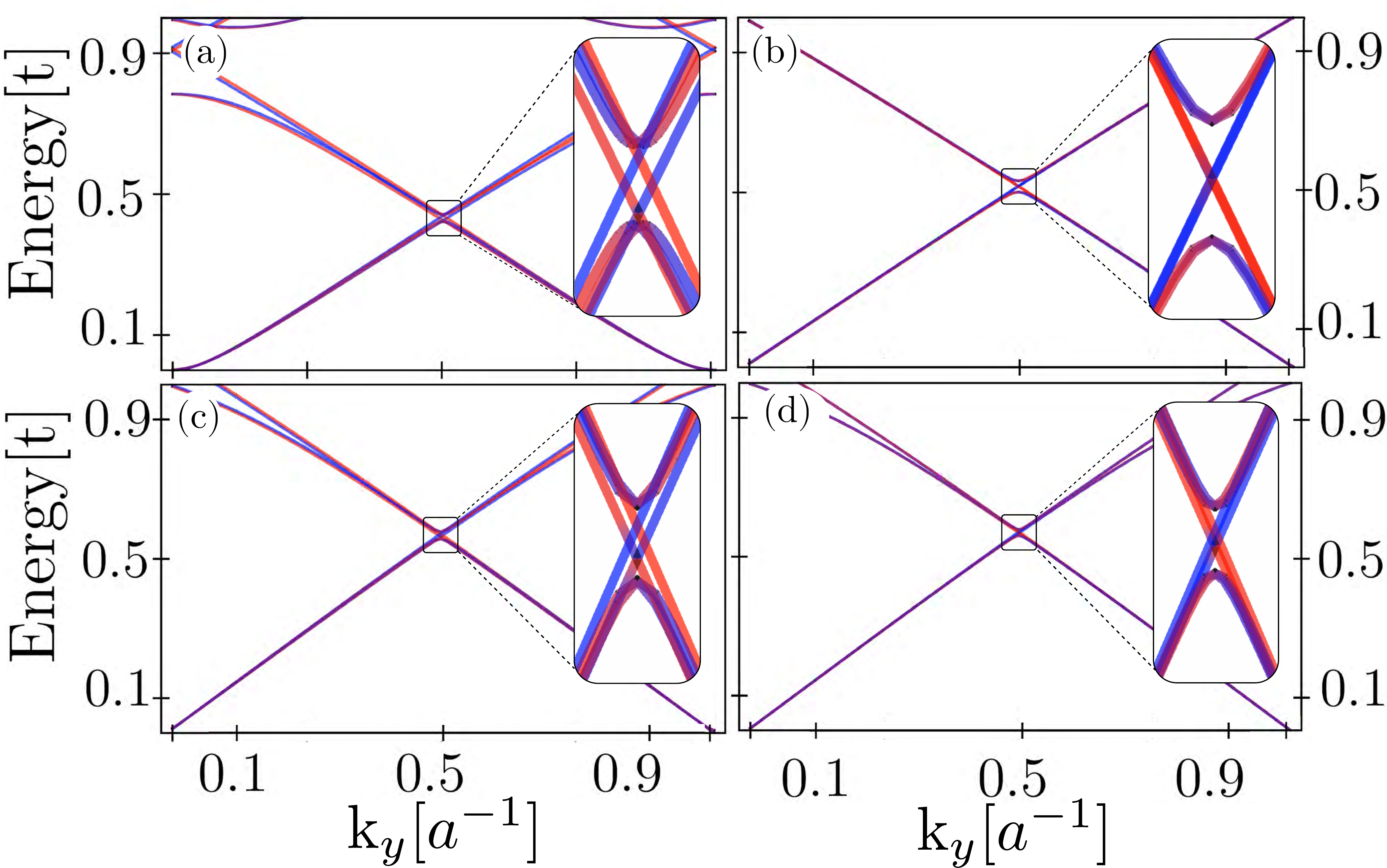}
		\caption{
		Energy spectra of  a metallic $N=3$ CNT for different models of implementing a DNA molecule wrapped at angle $ \varphi= \pi /6$. The effect of the DNA is incorporated by  adding to the bare Hamiltonian (Eqs.~\ref{eq:Hopping H}-\ref{eq:Sx}, with $\Gamma_1=10^{-4}t$ and $\Gamma_1=10^{-2}t$) different spatially-modulated terms on sites directly below the molecule. In (a) we added a potential with $U=0.5t$ underneath the molecule and $U=0$ otherwise. In (b) we modified the atomic SOC entering Eqs.~\ref{eq:Sy} and~\ref{eq:Sx} by taking $\Gamma_{1,2}\rightarrow10\Gamma_{1,2}$ underneath the molecule. In (c) the spin-independent hopping amplitudes are taken to be $t\rightarrow1.5t$  underneath the molecule. Panel (d) combines the modifications described  in (b) and (c). In all four cases, helicity-dependent gaps of similar magnitude open at equal momenta and energies.  }	\label{figureA1}
	\end{figure}

In the main text, we incorporated the molecule into the CNT Hamiltonian of Eqs.~(\ref{eq:Hopping H}-\ref{eq:Sx}) by adding a spatial modulation  to both spin-independent and spin-dependent hopping amplitudes. In that model, the dominant spin-dependent effects could be isolated by studying only the latter contribution, as we showed in the main text.  An alternative approach is to  introduce a spatially varying potential, $\delta\mathcal{H}=U\sum_{\mathbf{r},s}\left[A_{\mathbf{r},s}^{\dagger}A_{\mathbf{r},s}+B_{\mathbf{r},s}^{\dagger}B_{\mathbf{r},s}\right]$, where $U_{\mathbf{r}}\neq0$ is non-zero only for sites  directly below  the molecule. In Fig.~\ref{figureA1} we compare the electronic spectra of these different models for a metallic zig-zag CNT with $N=3$. We find that any of the implementations of the molecule, via a potential [panel (a)], through spin-dependent hopping [panel (b)] or through the spin-independent hopping [panel (c)] lead to similar gaps. The same holds for a combination of the latter two [panel (d)]. While the opening of helicity-dependent gaps is common to all of these, the details of the band structure may differ. In panels (b), the bands feature a two-fold degeneracy that is lifted in (a), (c) and (d). We note, however, that this non-universal property is \textit{spin independent}  and consequently does not affect spin filtering.


\section{Comparison between spatially modulated and Rashba SOC}\label{appendixB}

 \begin{figure}[b]
		\includegraphics[width=\linewidth]{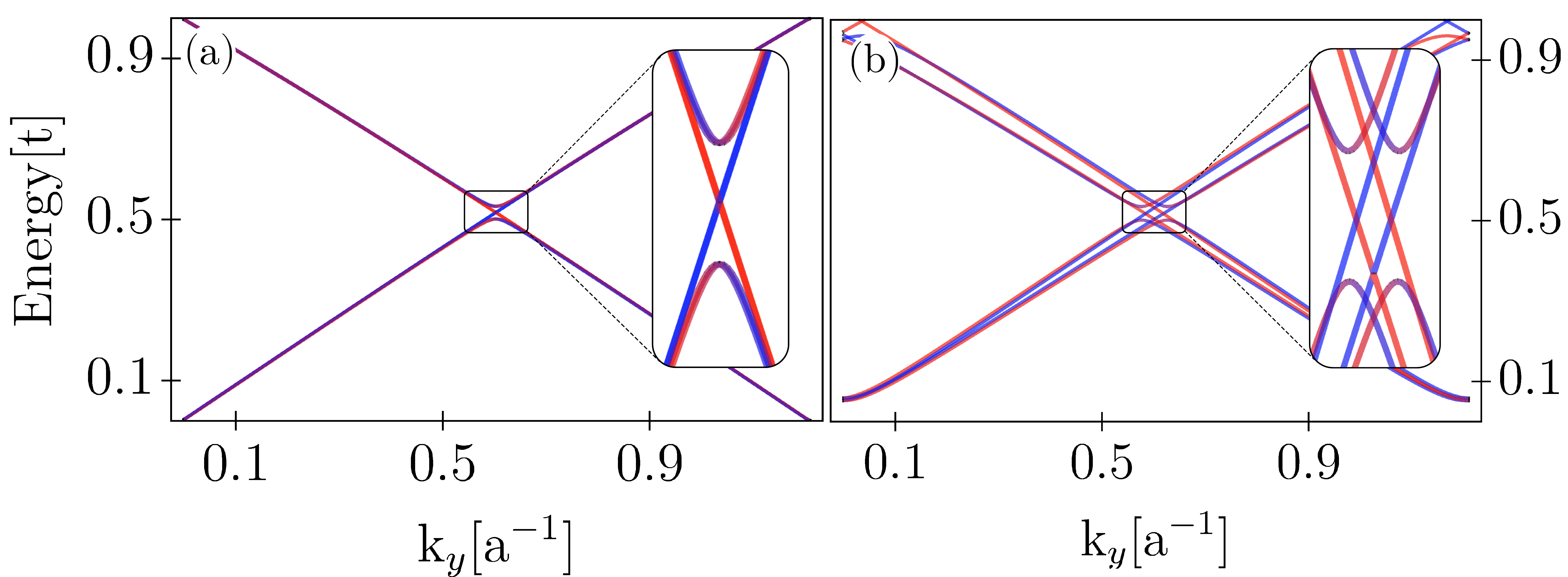}
		\caption{ 
		Comparison between the energy spectra of metallic $N=3$ CNTs wrapped by a DNA  molecule ($ \varphi= \pi /6$) in the presence of two types of spatial modulated SOC:  (a) Rashba and (b) curvature-induced. In (a) we perturb the bare graphene Hamiltonian, Eq.~\ref{eq:Hopping H}, with the Rashba term of Eq.~ \ref{eq:Rashba} (we set  $\Lambda=10^{-2}t$).  To facilitate comparison with the curvature-induced SOC we re-plot the corresponding spectrum of Fig.~\ref{figure3} in (b). As anticipated, both mechanisms give rise to  similar spectra; in particular, both open helicity-dependent gaps at the same energies and momenta.}
		\label{figureA2}
	\end{figure}
	
 	\begin{figure}[b]
		\includegraphics[width=1\linewidth]{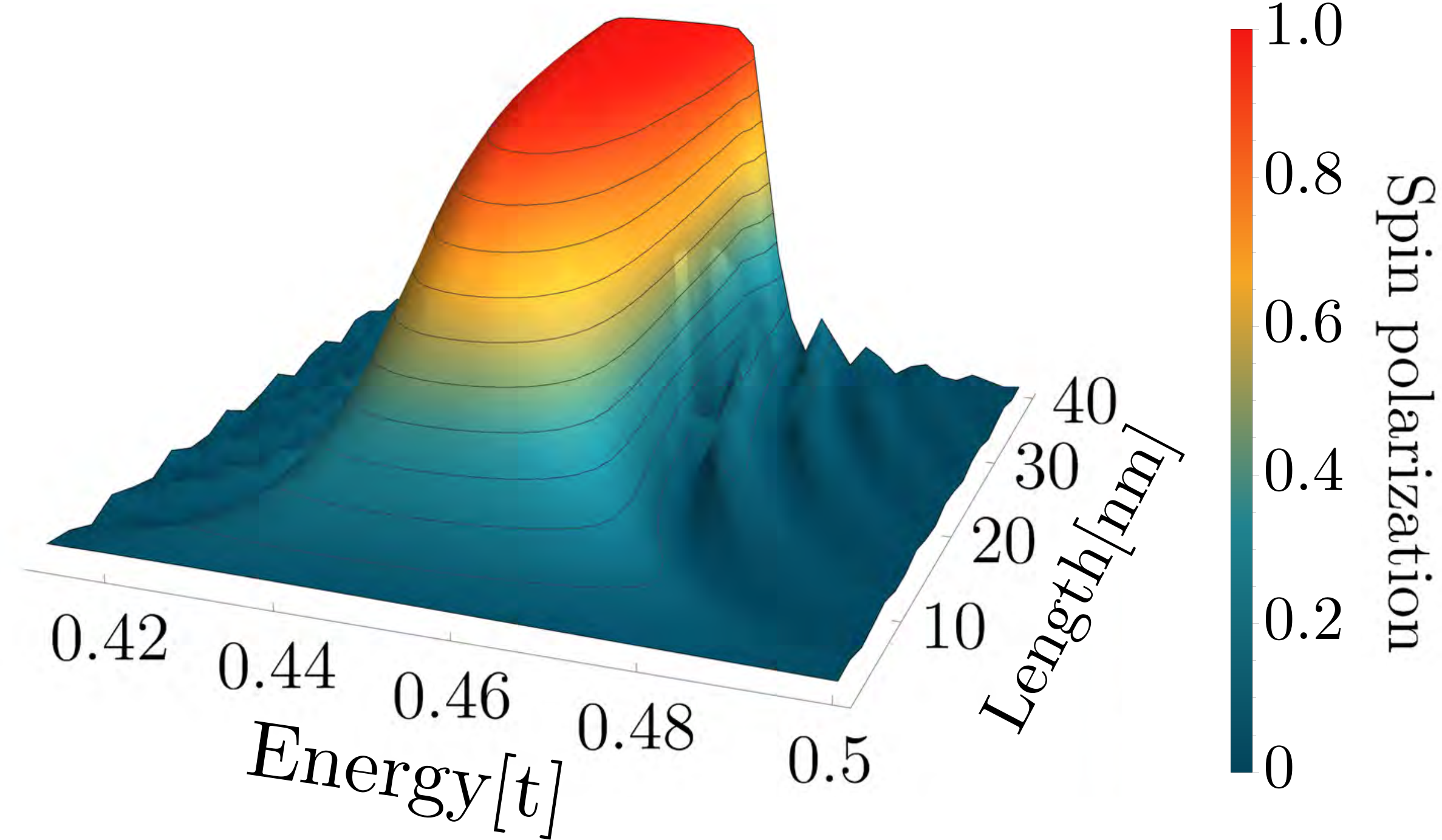}
		\caption{ Spin-polarization as a function of length for electrons transmitted through a semiconducting $N=5$ CNT wrapped by a right-handed DNA molecule ($ \varphi= \pi /6$). The spin-polarization saturates when the length exceeds $\approx 20 nm$, consistent with the estimate provided in the text. }
		\label{figureA4}
	\end{figure}

The model analyzed in the main text included only the $\pi$ orbitals, and neglected hybridization with the $\sigma$ orbitals due to the molecular potential. Allowing for such a mixing gives  rise to a Rashba SOC~\cite{Guinea2006,Diniz2012} 
\begin{align} \label{eq:Rashba}
\mathcal{H}_{\text{Rashba}}&=\sum_{\mathbf{r},i,s} \Lambda(\boldsymbol{r})\left[s (\hat{\delta}_i\cdot\hat{x})  A_{\boldsymbol{r}+\boldsymbol{\delta}_i,s}^{\dagger} B_{\boldsymbol{r},s}\right.\\\nonumber&\left.+
(\hat{\delta}_i\cdot\hat{y}) e^{-is(\theta_{\boldsymbol{r}}+\theta_{\boldsymbol{r}+\boldsymbol{\delta}_i})/{2}} A_{\boldsymbol{r}+\boldsymbol{\delta}_i,s}^{\dagger}B_{\boldsymbol{r},-s}+\text{h.c}\right],
\end{align}
where  $\Lambda(\boldsymbol{r})$ is non-zero only on sites directly below the molecule. The structure of this SOC resembles the curvature-induced one in Eqs.~\ref{eq:Sy} and~\ref{eq:Sx}; the main difference lies in the parameters $(\hat{\delta}_i\cdot\hat{x})$ and $(\hat{\delta}_i\cdot\hat{y})$, which are independent  of the curvature. Unsurprisingly, adding Eq.~\ref{eq:Rashba} to the bare graphene Hamiltonian of Eq.~\ref{eq:Hopping H} leads to helicity-dependent gaps in the energy spectrum, as illustrated in Fig.~\ref{figureA2}.


\section{Length-dependence of spin filtering}\label{appendixC}


The energy range over which spin polarization is significant is comparable to the magnitude of the partial gap, provided a sufficiently long segment of the CNT is wrapped with DNA.The distance where spin-polarization saturates is determined by $\xi=2v/\Delta$, where $\Delta$  is the size of the partial gap and $v$ is  the velocity of the unperturbed (ungapped) modes at these energies. In Fig.~\ref{figureA4} we show the spin-polarization as a function of energy and length for a semiconducting $N=5$ CNT. From its energy spectrum (Figs.~\ref{figure2} and~\ref{figure3}), we  extract the magnitude of the partial gap $\Delta\approx0.02t$ and the velocity $v\approx0.5 ta$. The resulting saturation length is $\xi\approx50a=20nm$, which is in agreement with our numerical result shown in Fig.~\ref{figureA4}.


\end{document}